\documentstyle[epsfig]{aipproc}

\def\msun{M$_\odot$}
\def\be7{$^{7}$Be}
\def\n13{$^{13}$N}
\def\o14{$^{14}$O}
\def\oo15{$^{15}$O}
\def\f17{$^{17}$F}
\def\ff18{$^{18}$F}
\def\na22{$^{22}$Na}
\def\al26{$^{26}$Al}

\begin{document}
\title{Predictions of gamma-ray emission from classical novae and their 
detectability by CGRO} 

\author{M. Hernanz$^1$, J. G\'omez-Gomar$^1$, J. Jos\'e$^2$ 
        and J. Isern$^1$}
\address{$^1$Institut d'Estudis Espacials de Catalunya (IEEC), 
         CSIC Research Unit\\
         Edifici Nexus-104, C/Gran Capita 2-4, 08034 Barcelona, SPAIN\\
$^2$Departament de F\'\i sica i Enginyeria Nuclear (UPC)\\
Avda. V\'\i ctor Balaguer s/n, 08800 Vilanova i la Geltr\'u, SPAIN}

\maketitle

\begin{abstract}
An implicit hydrodynamic code following the explosion of classical novae, 
from the accretion phase up to the final ejection of the envelope, has 
been coupled to a Monte-Carlo code able to simulate their gamma-ray emission.
Carbon-oxygen (CO) and oxygen-neon (ONe) novae have been studied and 
their gamma-ray spectra have been obtained, as well as the gamma-ray 
light curves for the important lines (e$^{-}$--e$^{+}$ annihilation line at 
511 keV, $^{7}$Be decay-line at 478 keV and $^{22}$Na decay-line at 1275 keV). 
The detectability of the emission by CGRO instruments has been analyzed. It 
is worth noticing that the $\gamma$-ray signature of a CO nova is different 
from that of an ONe one. In the CO case, the 478 keV line is very important, 
but lasts only for $\sim$2 months. In the ONe case, the 1275 keV line is the 
dominant one, lasting for $\sim$ 4 years. In both cases, the 511 keV line is 
the most intense line at the beginning, but its short duration ($\sim$2 days) 
makes it very difficult to be detected. 
It is shown that the negative results from the observations made by 
COMPTEL up to now are consistent with the theoretical predictions.
Predictions of the future detectability by the INTEGRAL mission are also made.
\end{abstract}
\section*{Introduction}
Nova explosions are caused by thermonuclear runaways on white dwarfs accreting 
hydrogen from a main sequence companion in a cataclysmic variable. The 
explosive burning of hydrogen on the top of a CO or an ONe degenerate core 
leads to the synthesis of some $\beta^+$ unstable nuclei, such as \n13, 
\o14, \oo15, \f17 and \ff18. These nuclei have short lifetimes ($\sim$1-2 
minutes for \o14, \oo15 and \f17, $\sim$15 min for \n13 and $\sim$3 hours 
for \ff18), and they emit a positron when they decay. These positrons 
annihilate with electrons leading to the emission of photons with energy 
less or equal to 511 keV. Furthermore, other 
medium- and long-lived radiocative nuclei are synthesized in classical novae:
\be7 ($\tau$=77days), whichs emits a photon of 478 keV after an electron 
capture, \na22 ($\tau$=3.75yr) and \al26 ($\tau$=1.04$\times 10^6$ yr), which 
experience a $\beta^+$-decay emitting photons of 1275 and 1809 keV, 
respectively. Thus, classical novae are potential $\gamma$-ray emitters, as 
was pointed out in previous works (Clayton \& Hoyle \cite{CH74}, Clayton 
\cite{Cl81}, Leising \& Clayton \cite{LC87}). There are also some previous 
works concerning nucleosynthesis in classical novae (see Politano et 
al \cite{Po95}, Prialnik \& Kovetz \cite{PK97} and references therein for 
nucleosynthesis in ONe and CO novae, respectively, 
as well as Hernanz et al \cite{He96}, Jos\'e et al \cite {Jo97}, 
Jos\'e \& Hernanz \cite{JH97a} and \cite{JH97b}). But to our knowledge there 
are no previous works which follow all the phases (from the accretion stage 
to the final explosion and mass ejection phases) and couple them to the study 
of the $\gamma$-ray emission of classical novae. 

In this paper we have used 
realistic profiles of densities, velocities and chemical compositions 
(obtained by means of a hydrodynamical code) to determine the production and 
transfer of $\gamma$-rays (by means of a Monte-Carlo code) during nova 
explosions. In this way, we are able to do a detailed analysis of the 
$\gamma$-ray emission of classical novae and predict their detectability by 
instruments onboard the CGRO (see our previous paper by Hernanz et al \cite 
{He97} for our first results and their relation to the future mission 
INTEGRAL). We are mainly concerned by the emission and potential detectability 
of individual novae, related to the decay of the medium-lived nuclei \be7 and 
\na22 (besides of the short-lived nuclei \n13 and \ff18).

\section*{$\gamma$-ray spectra and light curves of CO and ONe novae}
In table 1 we present the main properties of the ejecta of some of the 
most representative models we have computed (accretion rate 2$\times 
10^{-10}$\msun.yr$^{-1}$). 
There is an important difference 
between CO and ONe novae: CO novae are important producers of \be7 
(thus line emission at 478 keV is expected during some days), whereas ONe 
novae are important producers of \na22 (thus line emission at 1275 keV during 
the first years after the explosion is expected). This can clearly be seen 
by comparing the $\gamma$-ray spectra of a typical CO nova (CO1) and 
those of an ONe one (ONe2), shown in figure 1 for different epochs after 
the explosion.
In all cases a continuum component, mainly below 511 keV, appears as well as 
some lines (511 keV in all cases and 478 keV and 1275 keV in CO and ONe novae, 
respectively). Two phases can be distinguished in the evolution of all models. 
During the early expansion, the ejected envelope is optically thick and 
there is an important contribution of the continuum below 511 keV, related 
to the comptonization of 511 keV photons. Later on, when the envelope becomes 
optically thin, absorption and comptonization become negligible; therefore, 
the intensity of the lines is exclusively determined by the total mass of the 
radioactive nuclei \be7 (CO novae) and \na22 (ONe novae). The main properties 
of the lines are shown in table 2. It is worth noticing that fluxes are 
rather low, due to the small ejected masses (see table 1). 
This explains the null results 
of the observations of the 1275 keV emission by some novae made
with COMPTEL onboard CGRO (Iyudin et al 
\cite{Iy95}). Our predicted fluxes are fully compatible with the upper limits 
obtained in that work. It is also worth noticing the important fluxes 
associated with the 511 keV line that we predict for all models (see table 2). 
This flux should be detected by the current instruments onboard the CGRO, 
provided that novae are observed very early after their explosion. A 
retrospective analysis of the BATSE data would perhaps provide important clues 
to this issue (see Fishman et al \cite{Fi91}).
 
We want to mention that there is 
a lack of agreement between all the theoretical models of novae (including 
ours) and some observations of ejected masses of classical novae, although 
large uncertainties affect sometimes the determination of observed ejected 
masses. For instance, our model ONe1 fits quite well the observed abundances 
of the neon nova QU Vul 1984, but observations of this nova give an ejected 
mass that can reach $\sim10^{-3}$\msun (Saizar et al 1992), almost two 
orders of magnitude larger than the theoretical one. Thus the flux of the 
1275 keV line could be considerably larger, but no theoretical models are by 
now able to produce simultaneously such large ejected masses and neon in 
the ejecta.
\begin{table} 
\caption{Main properties of the ejecta one hour after peak temperature. 
Initial mass, total ejected mass and ejected mass of the most relevant 
radioactive nuclei are in \msun and the mean kinetic energy of the ejecta, 
$<\rm E_{\rm k}>$, is in erg.g$^{-1}$} 
\begin{tabular}{@{}lccccccc}
 Model & M$_{WD}$ 
       & M$_{\rm ejec}$  & $<\rm E_{\rm k}>$      & \be7             & 
         \n13            & \ff18                  & \na22            \\
  ONe1 & 1.15
       & 1.8~10$^{-5}$   & 3.1~10$^{16}$          & $\sim$ 0         & 
           5.5~10$^{-9}$ & 7.1~10$^{-8}$          &  9.8~10$^{-10}$   \\
  ONe2 & 1.25
       & 1.6~10$^{-5}$   & 3.3~10$^{16}$          & 1.2~10$^{-11}$   & 
           2.9~10$^{-8}$ & 6.7~10$^{-8}$          &  1.6~10$^{-9}$    \\
   CO1 & 0.8
       & 6.3~10$^{-5}$   &   8~10$^{15}$          & 7.8~10$^{-11}$   & 
           1.6~10$^{-7}$ & 1.7~10$^{-7}$          & $\sim$ 0          \\
   CO2 & 1.15
       & 1.4~10$^{-5}$   & 3.2~10$^{16}$          & 1.1~10$^{-10}$   & 
           1.3~10$^{-8}$ & 3.6~10$^{-8}$          & $\sim$ 0          \\
\end{tabular}
\end{table}

\begin{table} 
\caption
{Properties of the 511 keV line (columns 2 and 3) and of the 1275  and 478 keV 
lines (columns 5 and 6): time and flux (d=1kpc) at maximum.} 
\begin{tabular}{@{}lccccc}
Model  & t [hours] & Flux$_{511}$ [photons/s/cm$^{2}$]   
       & line     & t [days] & Flux [photons/s/cm$^{2}$] \\
ONe1   & 6        & 1.3~10$^{-2}$                          
       & 1275 keV & 7.5      & 3.8~10$^{-6}$ \\           
ONe2   & 5        & 1.6~10$^{-2}$                         
       & 1275 keV & 6.5      & 6.1~10$^{-6}$ \\      
CO1    & 7.5      & 1.4~10$^{-4}$ 
       &  478 keV & 13       & 1.4~10$^{-6}$ \\         
CO2    & 6.5      & 7.3~10$^{-3}$            
       &  478 keV & 5        & 2.6~10$^{-6}$ \\        
\end{tabular}
\end{table}
\begin{figure} 
\setlength{\unitlength}{1cm}
\begin{picture}(15,8)
\put(0,0){\makebox(7,8){\epsfxsize=7cm \epsfbox{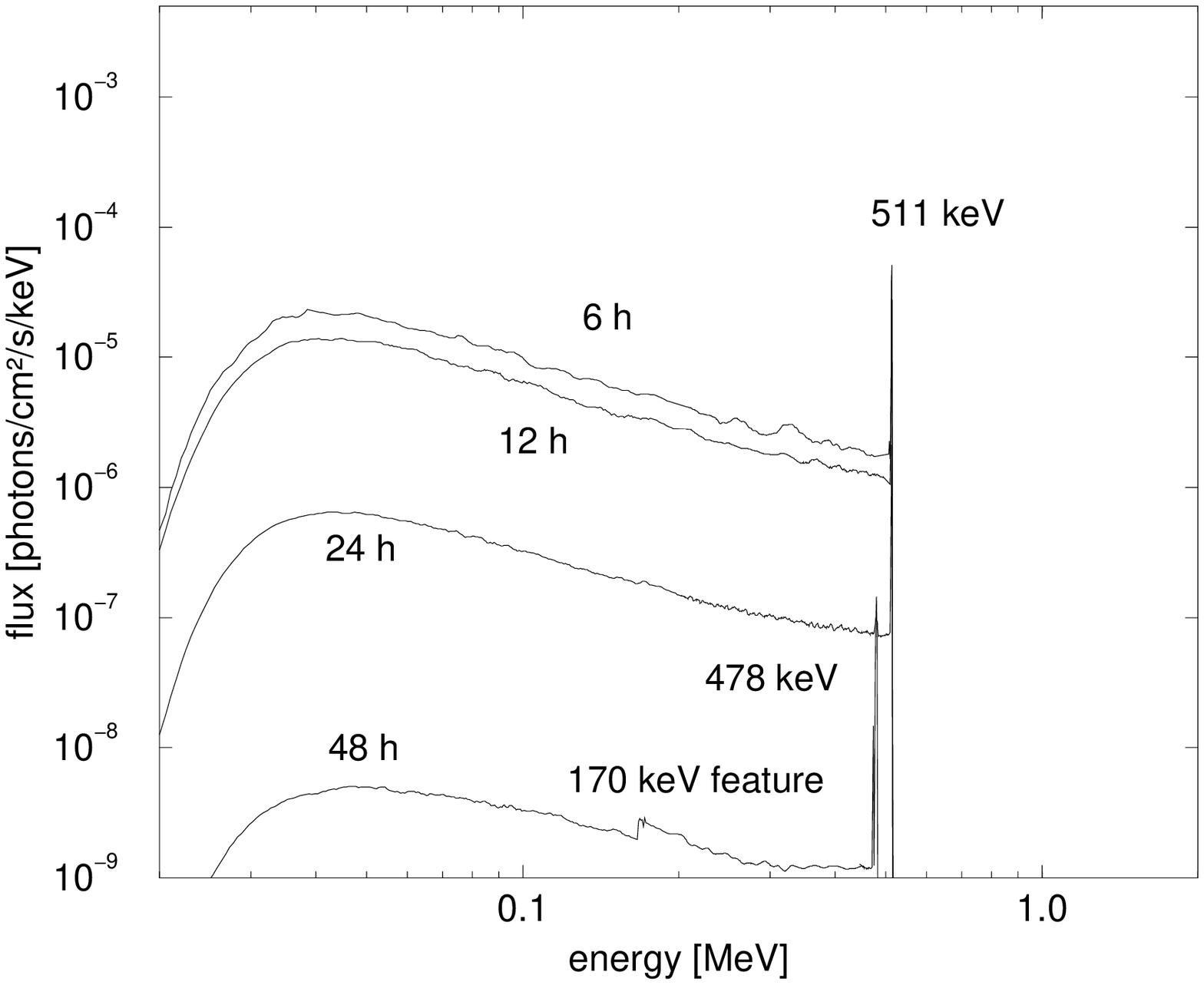}}}
\put(7.5,0){\makebox(7,8){\epsfxsize=7cm \epsfbox{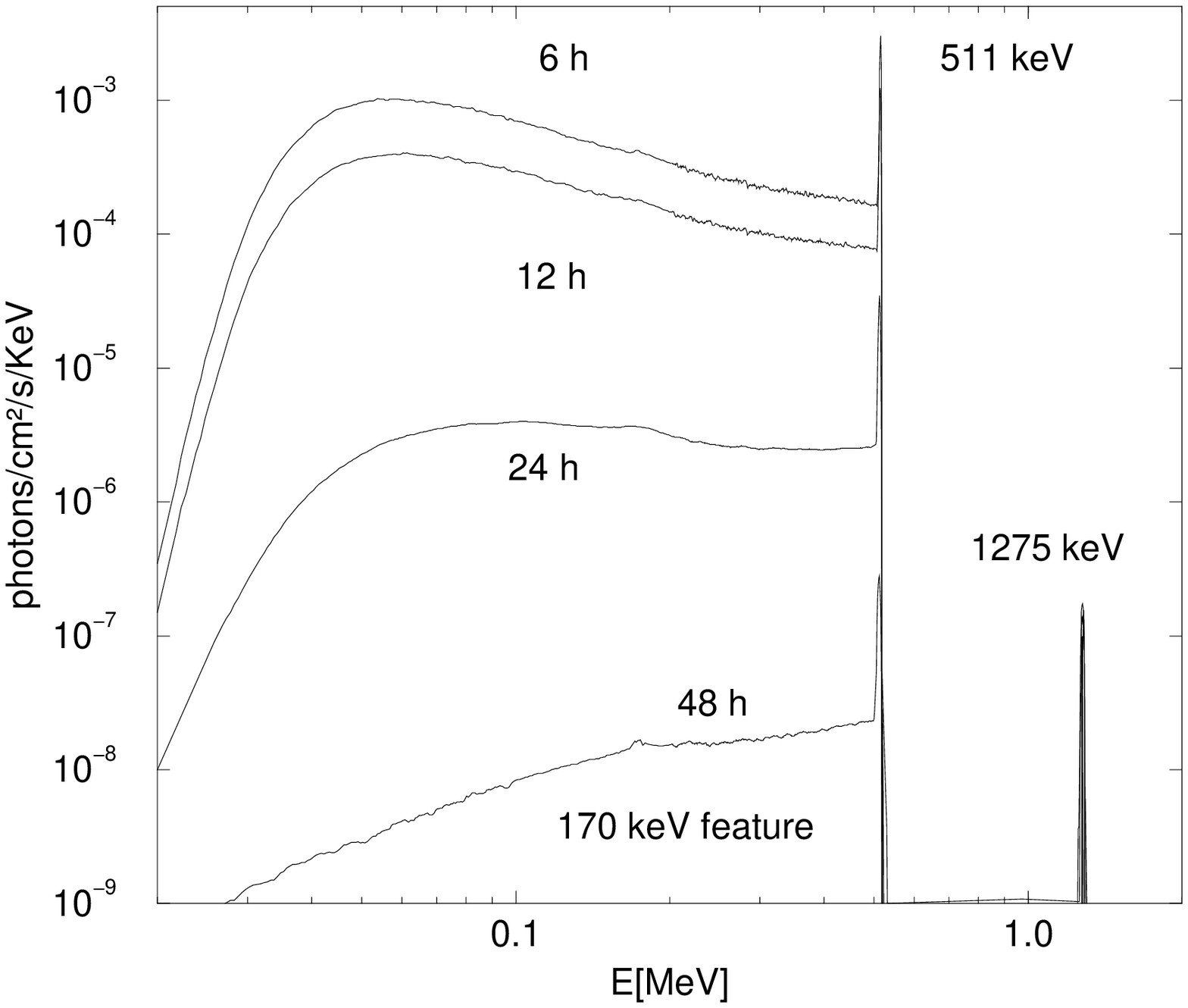}}}
\end{picture}
\vspace*{-1.5cm}
\caption{Evolution with time of the $\gamma$-ray spectrum of a CO nova of 
0.8 \msun (left) and an ONe nova of 1.25 \msun (right).}
\label{fig1}
\end{figure}

In figure 2 we show the light 
curves of the 478 and 1275 keV lines for CO and ONe models, respectively. For 
the 478 keV line two different phases can be distinguished: during the first 
$\sim$1.5 days, the line is completely dominated by the continuum generated 
by \n13 and \ff18 decays, whereas later on the line follows the typical light 
curve of the \be7 radioactive decay, with $\tau$=77 days. The 1275 keV line 
has a rise phase lasting $\sim$7days, followed by a decline related to the 
decay of \na22. The small fluxes obtained make it impossible to detect this 
line with the current CGRO instruments, unless a very close explosion 
occurs (the same as for the future SPI instrument onboard INTEGRAL).

Concerning the light curve of the 511 keV line for all the models, 
as several isotopes, with different decay timescales, contribute to this 
emission, its temporal evolution is somewhat complex. Besides one very 
early maximum appearing in the CO novae (related to \n13 decay), one later 
maximum appears in all cases at $\sim$ 6 hours (see table 2), which is related 
to \ff18 decay. The further development of the light curves is different in 
the CO and in the ONe cases: in the ONe case, as \na22 emits a e$^+$ when 
it decays, the emission at 511 keV lasts for a longer time (as the \na22 decay-
timescale is much longer than the \ff18 one). An early observation of a nova 
explosion, either of a CO or an ONe one, would provide a positive detection 
of the 511 keV line (an estimation of the detection distance of our novae 
by the future SPI instrument onboard INTEGRAL yields around 10 kpc for all 
of them except for the low-mass CO1 model).

\begin{figure} 
\setlength{\unitlength}{1cm}
\begin{picture}(15,8)
\put(0,0){\makebox(7,8){\epsfxsize=7cm \epsfbox{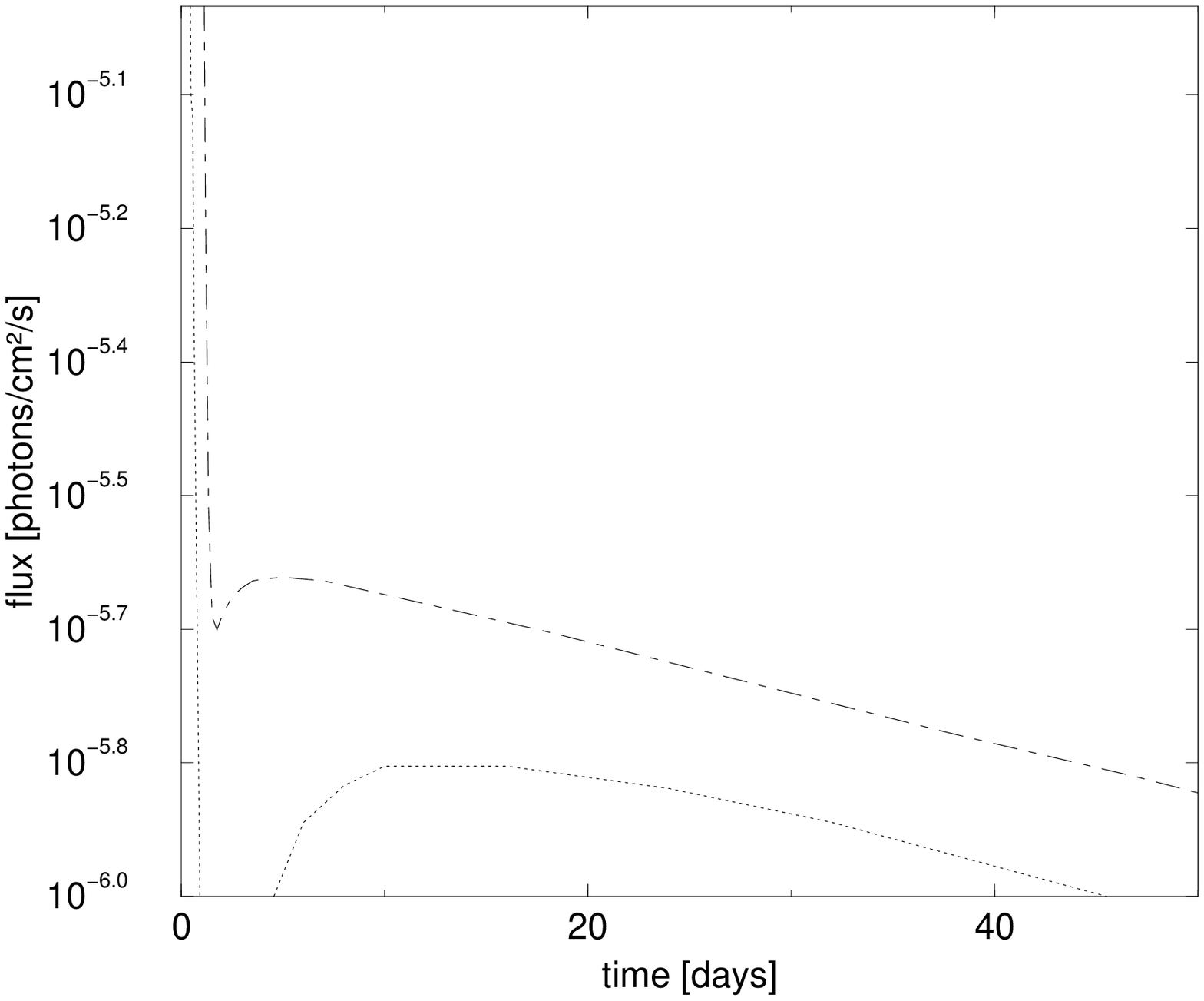}}}
\put(7.5,0){\makebox(7,8){\epsfxsize=7cm \epsfbox{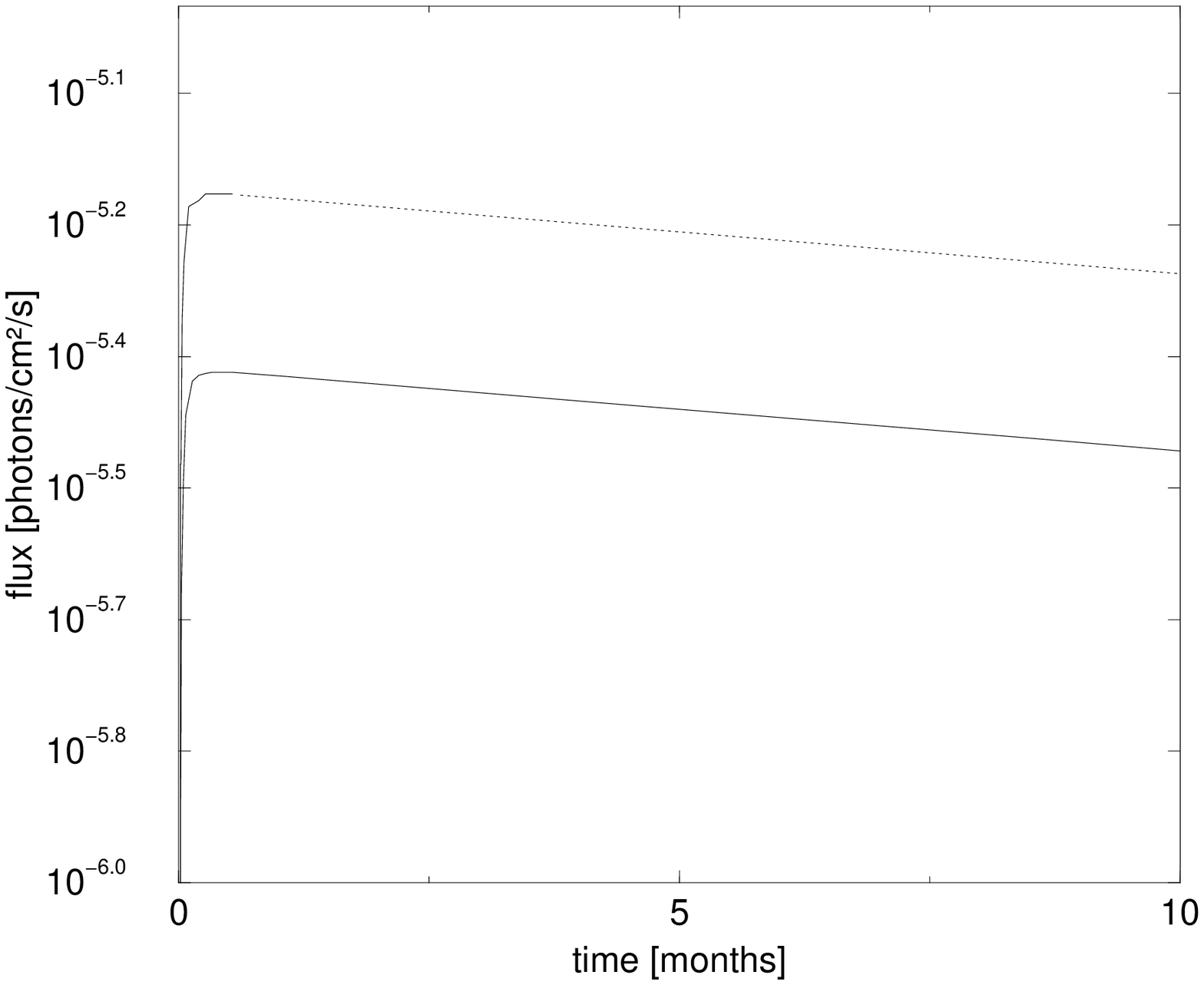}}}
\end{picture}
\vspace*{-1.5cm}
\caption{Left: Light curves of the 478 keV line (d=1 kpc) 
for CO novae (CO1: dotted, CO2: dot-dashed). Right: Light curves of the 
1275 keV line (d=1 kpc) for ONe novae (ONe1: solid, ONe2: dotted)}
\label{fig3}
\end{figure}

\section*{Conclusions}
We have developed a Monte-Carlo code in order to treat the production and 
transfer of $\gamma$-rays in nova envelopes, coupled to a hydrodynamical 
code that provides realistic profiles of all the relevant 
magnitudes. Thus a complete view of 
the nova explosion is obtained. The most relevant features of the $\gamma$-ray 
emission of classical novae are their intense 511 keV emission, lasting only 
some hours, the 478 keV emision in the case of CO novae, related to \be7 
decay, and the 1275 keV emission in the case of ONe novae, related to \na22 
decay. The 1275 keV emission is the most long-lived one, but the fluxes 
obtained are too low to be detected for novae at typical distances, thus 
explaining the negative detections of the novae observed up to now by COMPTEL.

\medskip
We thank for partial support the CICYT and DGICYT Projects ESP95-0091 and 
PB94-0827-C02-02.

\end{document}